\documentclass[12pt]{article}

\usepackage{graphicx,url,color}

\evensidemargin=\oddsidemargin

\title{\bf The incredibly strange story of Einstein's Nobel prize}
\author{Palash B. Pal\\
Department of Physics, University of Calcutta}
\date{December 2021}

\begin{document}

\maketitle

\begin{abstract}\noindent
  It is well-known that Einstein got the 1921 Nobel prize not for his
  theory of relativity, but for his theory of photoelectricity.  It is
  not that well-known that he did not get the prize in 1921.  Why not,
  and when did he get it?
\end{abstract}

On 25th October 2021, I got a frenetic email.  It was from an
ex-student of mine, who is now a professor at an undergraduate
college.  From the message, I learned that she is also the Head of the
Physics Deparment at the moment, and she had sent the message in her
capacity as the Head.  After these bits of information, the important
part of the message was contained in the next paragraph:

\begin{quote}

We are planning to organize a Webinar on the occasion of ``Centenary
year of the Nobel Prize to Einstein''. ...  We want to arrange the
webinar in the year 2021, date will be finalized according to your
convenience.

\end{quote}

To be sure, it is not clear how to detect freneticity in an email.
When one talks, the emotions come out through the pitch and the speed
of the words delivered.  Written words have no representation of
either of those qualities.  However, in this case, the signature was
unmistakable.  It was in the sentence ``We want to arrange the webinar
in the year 2021''.  Although it was not said explicitly, it would not
be unreasonable to guess that the reason for the hurry is the
occasion, as mentioned in the message itself: ``Centenary year of the
Nobel Prize to Einstein''.  Of course she was referring to the fact
that Einstein received the 1921 Nobel prize for Physics.

But there was no cause for hurry.  It can be debated whether 2021
should be the centenary year for the event of Einstein's Nobel prize,
or even whether there was an event at all.  In case you are wondering
what I mean by that statement, let me say, without beating around the
bush, that although Einstein got the 1921 Nobel Prize for Physics, he
did not get in 1921.

Nobel Prize for Physics is usually announced in October.  The
preparation for the choice starts the year before that.  In other
words, for the 2015 Nobel Prize, say, the first step was taken in
September 2014.  At that time, letters and messages were sent by a
five- or six-member Nobel Committee to a lot of luminaries, asking for
their suggestions for the next year's prize.  The deadline for
submitting suggestions is the end of the month of January, of the year
of the award.  A few months then go by while the Committee
deliberates, and also consults with experts outside the Committee.  In
June or July, the Committee prepares a report.  It includes the names
of the persons considered for the prize, some account of the
proceedings, and a recommendation for the recipient(s) based on
everything.  This report is submitted to the Royal Swedish Academy.
The Academy makes the final decision.  It may or may not uphold the
recommendation of the Committee.  The would-be recipient is announced
in October, and the prize is awarded in a ceremony at the Royal
Swedish Academy held on the 10th of December.
\begin{figure}
  \includegraphics[width=\textwidth]{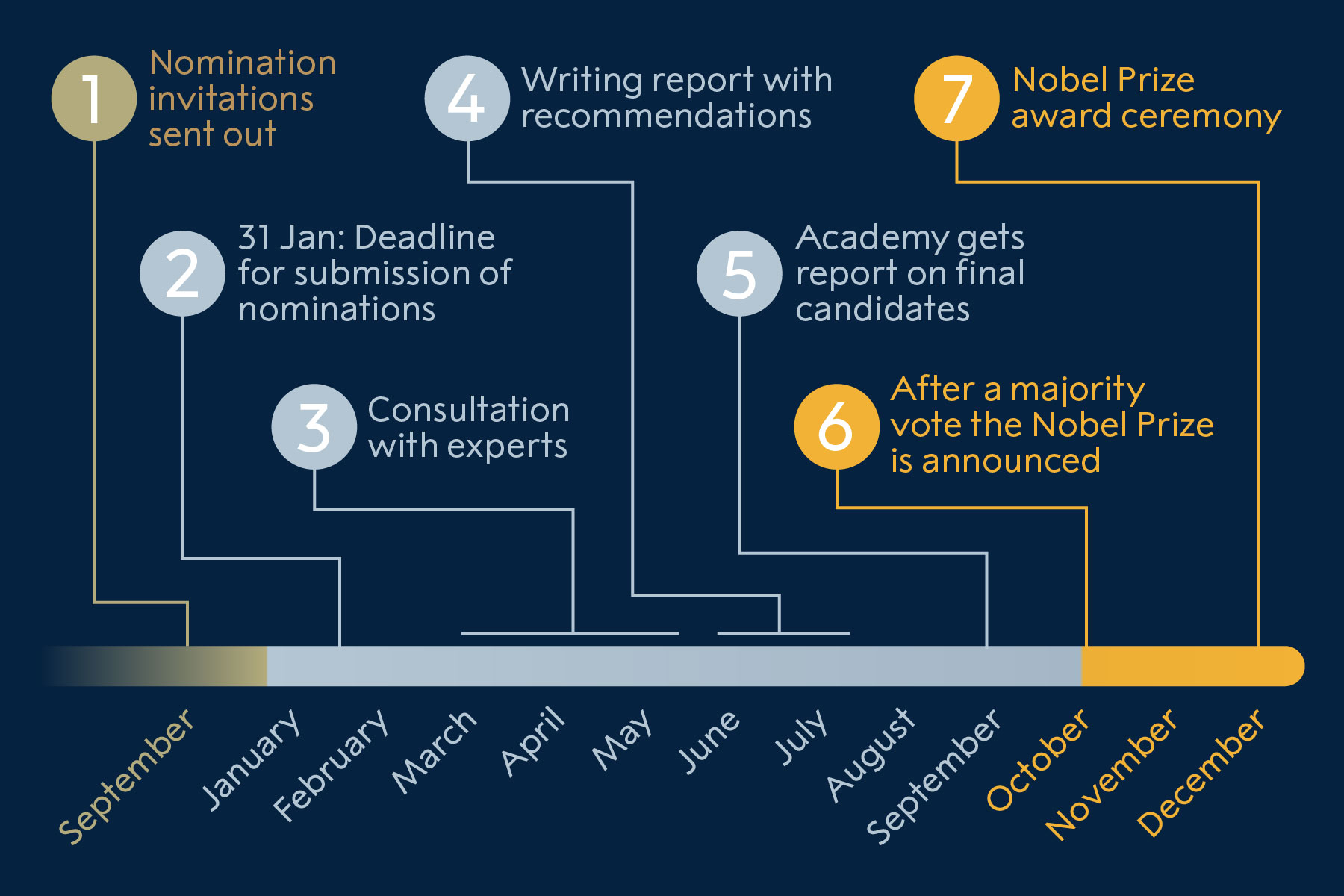}
  From: \url{https://www.nobelprize.org/nomination/physics/}
\end{figure}

Based on this timeline, one would conclude that Einstein, the
recipient of the 1921 Nobel Prize for Physics, came to know about the
decision in October 1921, and received the prize in December of the
same year.  No, that did not happen at all.  But we must go back a few
years before coming back to that part of the story.

Some of Einstein's phenomenal work appeared in 1905.  That year, in
fact, is often called the ``Annum mirabilis'', or the miracle year,
for Einstein.  He submitted at least five papers in that year, all of
which have become classic.  And, wonder of wonder, he did all the work
at a time when he did not have an academic position.  He applied for
quite a few positions earlier, and was denied.  Then, in 1902, he
accepted a job as technical expert third class at the patent office in
Bern.  During the Annum Mirabilis, he was still working there.

I will not get into a history of how his work was received in the
scientific community.  I will only mention that the recognition was
not instantaneous.  Even in 1907, when he applied for a position at
the University of Bern, his application was rejected.  He did not get
an academic position until 1908, and did not attend an academic
conference until 1909.  The position he obtained in 1908 was not a
permanent one.  His first permanent position was at the University of
Prague, where he moved in 1911.  Then, in 1912, he was offered a
professorship at ETH Z\"urich, which he accepted and moved back to
Switzerland.  Soon after, he was approached by people from Berlin,
with the idea of a membership at the Prussian Academy of Sciences.  He
accepted the position in December 1913, and moved to Berlin on the 6th
of April, 1914.

During all these turbulent years, his name started appearing in the
Nobel nominations.  In fact, with the exceptions of the years 1911 and
1915, he was nominated for the Nobel Physics prize every year from
1910 to 1922.  In 1910, there was just one person who proposed
Einstein's name.  The person was Wilhelm Ostwald, winner of the 1909
Chemistry Nobel, who, earlier in 1901, turned down Einstein's
application for an assistantship in his research group.  He was joined
by several others in 1912 and 1913, including some other Nobel
laureates.  All these nominations were for his theory of Special
Relativity.  Although some experimental confirmation of Einstein's
now-famous mass-energy relation started appearing in 1908, even in
1913 it was argued that the evidences were not compelling enough to
merit a prize for the theory.  Some other confirmation of the theory
appeared later, 1916 onward.  But by then, Einstein had published his
General Theory of Relativity (GTR), and the tables had turned.  In
fact, in 1916, he was nominated for his GTR, and also for his work on
Brownian motion which demonstrated the existence of molecules.  There
was only one nomination in that year.

Let's jump cut to 1921.  By then, Einstein had become a household
name.  This is not an exaggeration.  Indeed his fame spread much
beyond the circle of physicists, even to lay people not directly
involved with science, because of the experiments performed by Arthur
Eddington in 1919, which proved that the path of light is bent by
gravitation, in agreement with the prediction made in Einstein's GTR.
In 1920, his nominations talked mainly about the GTR, as expected.

For the 1921 prize, there was a strong note from Max Planck, who had
won the prize for 1918, nominating Einstein for GTR.  It was a repeat
of the request he had made the earlier year.  Carl Oseen, a
theoretical physicist from the University of Uppsala, proposed
Einstein's name for the photoelectric effect.  The committee requested
its member Allvar Gullstrand to prepare a report on the theory of
relativity, and another member, Svante Arrhenius, a report of the
photoelectric effect.  They were both celebrated sceintists.
Gullstrand was a professor of opthalmology who had worked on
geometrical optics and won the 1911 prize for Physiology and Medicine
for his contributions to the functioning of the human eye.  His report
on relativity was highly critical of the theory.  He surmised that the
effects of relativity would be unmeasurable, ``so small that in
general they lie below the limits of experimental error''.  On the
other side, Arrhenius, the 1903 Nobel laureate for Chemistry who is
sometimes called the ``father of Physical Chemistry'', said that since
Planck had been awarded the Nobel for quantum theory as recently as in
1918, another prize should not go in the same direction so soon.  So,
near the end of the year 1921, the Nobel Prize committee decided that
they could not find anyone suitable for the prize for that year.  The
year of this writing, 2021, marks the centenary of that decision.

In 1922, Einstein's name was proposed again.  Meanwhile, the campaign
for him had increased in strength.  Maurice Brillouin (not the person
who is responsible for the idea of ``Brillouin zone'') wrote,
``Imagine for a moment what the general opinion will be fifty years
from now if the name Einstein does not appear on the list of Nobel
laureates.''  Oseen repeated his nomination for the photoelectric
effect.  Planck proposed to give the overdue 1921 prize to Einstein
and the 1922 prize to Niels Bohr.  This was indeed possible, because
according to the Nobel Foundation's statutes, when no acceptable
candidate can be found, the Nobel Prize can be reserved until the
following year.

The Committee again asked Gullstrand for a report on relativity, and
Gullstrand reiterated what he had written the year before.  Oseen was
asked as well, to produce a report on photoelectric effect, and he
gave an excellent account of Einstein's revolutionary contribution to
its theory.  The Committee recommended Einstein for the 1921 prize,
and Bohr for the 1922 prize, just as Planck had suggested.  The
Swedish Academy upheld the decision.  Accordingly, a telegram was
delivered to Einstein's address in Berlin on the 10th of November
1922, informing him of the good news.

\bigskip

Now you might have started to think that the rest of the story would
be obvious: that Einstein went to Stockholm in the December of 1922
and received his prize, the Nobel Physics prize for the year 1921.
Wrong.  That's not what had happened.

Einstein was not home when the telegram arrived.  He and his wife Elsa
were on board a ship to Japan.  He would not be back until March
1923.  So, obviously he did not go to Stockholm to receive his prize
in December 1922.

Certainly Einstein would not have looked at it as a lost opportunity.
In fact, during the last few years, he was really quite sure that he
would receive the Nobel Prize at some point.  When he got divorced
from his first wife Mileva Mari\v{c} in January 1919, he promised her
that he would give her the entire Nobel Prize money, whenever he would
get it.  So, clearly, the news of the Nobel Prize did not come as a
surprise to Einstein.

Indeed, in September of 1922, the 1914 Nobel Physics winner Max von
Laue wrote a letter to Einstein, telling him that it would be
``desirable for you to be present in Europe in December.''  Laue knew
about Einstein's plan for visiting Japan, and he advised Einstein not
to go.  Of course he did not mention the Nobel Prize, but the hint was
rather obvious.  Einstein ignored Laue's suggestion.  It may also be
true that he wanted badly to get out of Germany at that time.  On 24th
June 1922, Walter Rathenau, a politician and Einstein's friend, was
assassinated.  There were attempts at the lives of several other
people, which were clear signs of rising anti-semitism in Germany.

Anyway, the news must have reached Einstein during his journey.  It is
not clear when that happened.  There was no mention of the arrival of
the news in his diary.

It was clear, much before December, that Einstein would not be around
to accept his prize.  The question then arose: who would receive it on
his behalf?  Usually, when a recipient is absent for some reason, the
person who represents him or her at the ceremony is the Ambassador of
the recipient's native country to Sweden, residing in Stockholm.  So
now a battle of sorts ensued: the bone of contention being, what was
Einstein's native country?

Einstein was born in Germany in 1879 in the southern town of Ulm, so
he must have been a German citizen by birth.  But he moved with his
family: first to Italy in 1894, and then to Switzerland in 1895.  In
1896, he paid 3 marks to obtain a document that proclaimed that he was
not a German citizen anymore.  He remained stateless for the next five
years, until he got the Swiss citizenship in 1901.  At the time when
the Nobel prize was announced for him, he was traveling with his Swiss
passport.  Clearly, then, it should have been the Swiss Ambassador to
Stockholm who should accept the prize for him!

But this is a story where nothing goes according to the expectation.
The information about Einstein's passport was known to the German
Foreign Office, who passed it on to the German Ambassador in Sweden,
Rudolph Nadolny.  Nadolny refused to believe it.  He cannot be blamed
for that.  In 1922, Einstein was in Berlin, a member of the Prussian
Academy of Sciences.  That position was available only to German
citizens.  When Nadolny telegraphed to the Academy, inquiring about
the citizenship, the Academy replied what it knew and believed: that
Einstein was indeed a German.  Nadolny showed the reply to the Swiss
Ambassador.  He was surprised, probably unhappy, but he accepted the
situation.  He thought that probably, since Einstein had been living
in Berlin for a while, he considered himself to be a German again.
Nadolny graciously stated that Einstein's Swiss connection be duly
mentioned in public statements of any kind, and that the Swiss
Ambassador be also invited to the ceremony and to the banquet
afterwards.  In any case, it was Nadolny who accepted the prize on
behalf on Einstein in December 1922, and in the Nobel records he
appeared as a German.

The matter did not end there.  The German Ministry of Science asked
the Berlin Academy to clear up the citizenship issue.  The Academy
sent its report on 23rd January 1923.  They said that, since the job
at the Academy required a German citizenship and Einstein had accepted
the job, it could be concluded that he was a German.  Remember that
Einstein was away in Japan when all these things were going on.  When
he returned in March 1923, he was asked to give his view on the
matter.  His letter, dated 24th March 1923, contained the following
lines:
\begin{quote}
  When my appointment to the Academy was being considered, my
  colleague Haber informed me that my appointment would result in my
  becoming a Prussian citizen.  As I attached importance to retaining
  my nationality, I made acceptance of a possible appointment
  dependent on this, a stipulation which was agreed to.
\end{quote}
It was clear that Einstein remained a Swiss citizen at heart.  So,
when it came to the issue of handing over the Nobel prize to him, he
preferred that it be done by the Swiss Ambassador to Germany.  In
fact, on 6th April 1923, Einstein's stepdaughter Ilse wrote a letter
to the Nobel Foundation, informing that Professor Einstein would
appreciate if the medal and diploma could be sent to him in Berlin,
and added that it is preferred that this is done through diplomatic
channels, and the ``Swiss Embassy should be considered, since
Professor Einstein is a Swiss citizen.''  Finally, the Swedish
Ambassordor to Germany, Baron Ramel, handed the prize over to Einstein
in Berlin.

To summarize, if we were to celebrate the centenary of the Nobel prize
announcement or the formal presentation of the award, that would be in
2022.  If we want to celebrate the centenary of Einstein actually
receiving the prize, that should happen in 2023.

\bigskip

There are a few things about the prize which have not been discussed
yet.  First, what happened to the money?  As I have already mentioned,
Einstein promised his first wife Mileva, at the time of their divorce
in January 1919, that she would receive the entire Nobel prize money.
Einstein kept up to his promise.  Soon after he got the money, in
1923, he transferred the entire amount to the bank account of Mileva.
Mileva bought a house in Z\"urich.  Albert Einstein himself did not
enjoy a penny of his Nobel prize.

The other pending thing is about the Nobel lecture.  A Nobel prize
winner is supposed to give a Nobel lecture at the time when the prize
is conferred, on the work cited in the award.  In Einstein's case, the
winner was absent at the ceremony in December of 1922.  After he came
back to Europe in March 1923, he received a letter from Svante
Arrhenius, Director of the Nobel Institute and an esteemed member of
the Nobel Committee, requesting him to visit Sweden at a time of his
conveience, and telling him that he need not wait till the next Nobel
ceremony which would be held in December 1923.  So Einstein organized
a trip to Sweden.  On 11th July 1923, he gave a lecture: not in
Stockholm in front of the Royal Swedish Academy, but to the Nordic
Assembly of Naturalists at G\"oteborg.  The audience consisted of
about 2000 people, including the King of Sweden.  Because of the lack
of anything better, this lecture is considered to be his Nobel lecture
ever since, not only unofficially but even by the Nobel Prize
Foundation, and can be obtained on their website, and printed in
authorized collections of Nobel Prize lectures.

But, unlike usual Nobel lectures, this was not a lecture of the work
that was cited in the award.  The citation said that Einstein was
receiving his prize ``for his services to Theoretical Physics, and
especially for his discovery of the law of the photoelectric effect.''
There was no mention of relativity.  There are good reasons to believe
that the subject was deliberately omitted, since in his nomination
history the reports on relativity were not favorable.  However, by an
irony of fate, when Arrhenius invited him to Sweden, he left the
choice of the topic to the speaker, but added that something on his
relativity theory would be a good choice.  So Einstein gave his
G\"oteborg lecture on ``Fundamental ideas and problems of the theory
of relativity''.  There was absolutely no reference to his own work on
photoelectric effect: not even once.  The Nobel Foundation had to
accept it as the overdue Nobel lecture.  All they could do is to add a
comment: ``The Lecture was not delivered on the occasion of the Nobel
Prize award, and did not, therefore, concern the discovery of the
photoelectric effect.''  This comment appears as a footnote in the
Foundation's website.

Was relativity totally overlooked in Einstein's Nobel prize?
Certainly the prize was not given for relativity.  Even if one accepts
the argument, however lame it might be, that there was not enough
experimental evidence to support the theory in 1921 or 1922, the
Swedish Academy could have considered him for a second Nobel for the
Theory of relativity at some later time.  They did not.  It is a pity
that there was no member in the Committee who could evaluate
relativity at that time.  The reports submitted on relativity were
negative, and certainly the prize could not be given on the basis of
them.  The Nobel Committe probably began feeling that it would be an
embarrassment if Einstein hadn't received the Nobel prize.  So, to
quote Einstein's biographer Abraham Pais, ``Oseen's proposal to give
the award  for photoelectricity must have come as a relief of
conflicting pressures.'' 

However, it has to be said that relativity was not completely omitted
in the entire event.  The presentation speech for the 1921 Nobel
Physics prize was made by Svante Arrhenius.  He started his speech
like this:
\begin{quote}
  There is probably no physicist living today whose name has become so
  widely known as that of Albert Einstein. Most discussion centres on
  his theory of relativity. This pertains essentially to epistemology
  and has therefore been the subject of lively debate in philosophical
  circles. It will be no secret that the famous philosopher Bergson in
  Paris has challenged this theory, while other philosophers have
  acclaimed it wholeheartedly. The theory in question also has
  astrophysical implications which are being rigorously examined at
  the present time. 

\end{quote}

So, Arrhenius admired relativity and recognized its importance.  But
he thought that it was essentially ``epistemology''.  The word,
according to the Merriam-Webster dictionary, means ``the study or a
theory of the nature and grounds of knowledge especially with
reference to its limits and validity''.

Maybe it was not possible to anticipate what role the theory would
play in the coming decades, in the discovery of new elementary
particles, in the understanding the behavior of heavy atoms, and in
the birth and development of the theory of the universe as a whole.
But from the presentation speech, it seems that the people in Nobel
Foundation did not even realize that relativity was a theory of
Physics.  They brought up Bergson's philosophical arguments in the
context.  Einstein liked and respected Bergson as a person.  About his
philosophy, he apparently commented, ``Gott verzeih ihm'' (God forgive
him).

Was Einstein hurt, for no mention of relativity, then or ever again,
in the circle of Nobel awards?  Probably he did not care.  The fame
came to him anyway.  The money did not come to him.  As for the prize
itself, or in general about prizes, he was probably not a big
enthusiast.  A few incidents might throw light on his attitude.

The Planck medal is the highest award of the German Physical Society.
It was instituted in 1929, and Einstein was the recipient of the
inaugural year.  The day of the award, he did some work in the morning
and went to the house of his doctor friend, Janos Plesch, for lunch.
After lunch, he fell asleep on a couch.  He got up at four.  The
ceremony was supposed to begin at five.  Suddenly he realized that he
might be asked to speak at the occasion.  So he sat down at Plesch's
table, and grabbed the nearest piece of paper, which happened to be a
bootmaker's bill.  He scribbled on it for twenty minutes.  Half an
hour later, when Planck awarded the medal to him, he said, in his
acceptance speech, that he knew that he would be overwhelmed after
receiving the prize and would be at a loss for words, so he had
written down his speech, and would read it out.  He pulled the shoe
bill from his waistcoat pocket and started reading.  After the speech,
Plesch told him that he needed the bill back.  Einstein reached in his
pocket, pulled the bill and the medal that was wrapped in it, and gave
the whole thing to Plesch.  Plesch wrote, ``He never took it out, and
never looked at it again.''

In 1932, he was invited to become a member of a science academy.  A
form was sent to him, to be filled out.  It contained nine questions,
including birth details, education, publications and major scientific
contributions.  Einstein completed the form the submitted it.  There
was no mention of his Nobel prize.  Certainly he considered it to be
irrelevant to the answers of the questions asked.

Probably there is nothing to be surprised about, at least in the light
of the other things that we summarized here.  It is said that truth is
stranger than fiction.  Certainly the entire story appears almost
unreal, or maybe magic-real.  That is why I could not help using a
Marquezian title to this article, mimicking ``The incredible and sad
tale of innocent Erendira and her cruel grandmother''.

\renewcommand{\refname}{\normalsize Material obtained from:}


\begin{thebibliography}{[99]}

\bibitem{pais} Abraham Pais, ``Subtle is the Lord'' (1982).

\bibitem{hoffman} Banesh Hoffmann, ``Albert Einstein creator and
  rebel'' (1972).

\bibitem{clark} Ronald W.\ Clark, ``Einstein: the life and times''
  (1971).

\bibitem{foundation} \url{https:// www . nobelprize . org/}
  

\bibitem{prizemoney} Barbara Wolff, \url{https://www . einstein-
 website . de / z_information/ nobelpreisgeld . html}

\bibitem{gg} Gautam Gangopadhyay, \url{https:// gautamgangopadhyay
 . blogspot . com/ 2021/ 11/ blog-post_19 . html}

\bibitem{pbp} Palash Baran Pal, ``Einstein-er uttoradhikar'' (2011).
  
  


\end{thebibliography}
\end{document}